\documentclass[sigconf]{acmart}

\def\BibTeX{{\rm B\kern-.05em{\sc i\kern-.025em b}\kern-.08emT\kern-.1667em\lower.7ex\hbox{E}\kern-.125emX}}
    
\copyrightyear{2019}
\acmYear{2019}
\setcopyright{acmlicensed}
\acmConference[KDD-ADF-2019]{2nd KDD Workshop on Anomaly Detection in Finance}{Aug 05, 2019}{Anchorage, AK}

\usepackage{multirow}
\usepackage{multicol}

\begin{document}

%
\title[Automating Data Monitoring: Detecting Structural Breaks in Time Series Data Using BMDL]{Automating Data Monitoring: Detecting Structural Breaks in Time Series Data Using Bayesian Minimum Description Length}

%
\author{Yingbo Li}
\email{yingbo.li@capitalone.com}
\affiliation{%
  \institution{Capital One}
  \streetaddress{8000 Dominion Pkwy}
  \city{Plano}
  \state{TX}
  \postcode{75024}
}

\author{Robert Cezeaux}
\email{robert.cezeaux@capitalone.com}
\affiliation{%
  \institution{Capital One}
  \streetaddress{8000 Dominion Pkwy}
  \city{Plano}
  \state{TX}
  \postcode{75024}
}

\author{Di Yu}
\email{di.yu@capitalone.com}
\affiliation{%
  \institution{Capital One}
  \streetaddress{8000 Dominion Pkwy}
  \city{Plano}
  \state{TX}
  \postcode{75024}
}

%
\renewcommand{\shortauthors}{Li, et al.}

%
\begin{abstract}
In modern business modeling and analytics, data monitoring plays a critical role. 
Nowadays, sophisticated models often rely on hundreds or even thousands of input variables.
Over time, structural changes such as abrupt level shifts or trend slope changes may occur among some of these variables, 
likely due to changes in economy or government policies. 
As a part of data monitoring, it is important to identify these changepoints, 
in terms of which variables exhibit such changes,
and what time locations do the changepoints occur. 
Being alerted about the changepoints can help modelers decide if models need modification or rebuilds,
while ignoring them may increase risks of model degrading. 

Simple process control rules often flag too many false alarms because
regular seasonal fluctuations or steady upward or downward trends usually trigger alerts.
To reduce potential false alarms, we create a novel statistical method based on the 
Bayesian Minimum Description Length (BMDL) framework 
to perform multiple changepoint detection.  
Our method is capable of detecting all structural breaks occurred in the past, 
and  automatically handling data with or without seasonality and/or autocorrelation. 
It is implemented with computation algorithms such as Markov chain Monte Carlo (MCMC),
and can be applied to all variables in parallel. 

As an explainable anomaly detection tool, our changepoint detection method 
not only triggers alerts, but provides useful information about the structural breaks, 
such as the times of changepoints, and estimation of mean levels and linear slopes before and after
the changepoints. This makes future business analysis and evaluation on the 
structural breaks easier.
\end{abstract}

%
%

%

%

%
\maketitle

\section{Introduction}

Modern business models often have hundreds of variables.
Over time, some of the variables may experience structural changes
in mean levels and/or linear slopes,
likely due to changes in economy or government policies.
Hence, data monitoring plays a critical role to identify these changepoints.
With knowledge of the changepoints, modelers can take actions
to modify models and thus minimize risks of model degrading.

One common approach of data monitoring is to apply process control tools, such as
Shewhart control chart rules \citep{Shewhart_1931}. 
For example, an observation being beyond three sigma (standard deviation) 
away from the centerline triggers an alert. These tools, 
while being successful for quality control purposes, 
are hardly as successful in business and financial applications.
They tend to be too sensitive and thus trigger too many alerts
for modelers to conduct further investigation. 
This is because for business applications, 
even without a structural break, a time series variable often presents
trend, seasonality, and autocorrelation. However, control chart rules 
usually fail to take these into account. 
Rather, any of these can make the variable to be incorrectly flagged as an anomaly. 

In this paper, we introduce a novel model-based changepoint detection  approach,
which is flexible, explainable, easy to automate, and most importantly, 
effective in reducing false positives. 
We follow the Bayesian Minimum Description Length (BMDL) framework in \citet{Li_etal_2019},
and extend the method there to handle more flexible data monitoring tasks.
Our method automatically detects not only 
the optimal combinations of changepoint locations in the past,
but also whether the time series data has seasonality and/or autocorrelation.
To aid explainable analysis, our method provides information on
how the changepoints affect the data, by creating
estimations of mean levels and linear slopes before and after each changepoint.
It can be efficiently implemented with stochastic search algorithms such as 
Markov chain Monte Carlo (MCMC). 
Designed for univariate time series data, our method enables scalability 
as it can be applied to hundred of variables at the same time, in parallel. 
Since the BMDL framework is essentially penalization
based model selection, our method acts relatively conservative in
flagging changepoints. Therefore, it substantially reduces potential false positives. 

The rest of the paper is organized as follows. 
In Section \ref{sec:methods},
we first introduce the concept of a multiple changepoint configuration,
then show how the sampling distribution accommodates
regime-wise linear segments, a seasonal mean cycle, and autoregressive errors.
Next, we give high level details of the BMDL derivation, and
last briefly discuss computation.
In Section \ref{sec:results}, we give two examples
to demonstrate the performance of our BMDL method.
The first example is to detect historical changepoints in a publicly available 
dataset on average earnings in metropolitan areas.
The second example is to evaluate data monitoring false positive and
true positive rates using simulation data and compare it with Shewhart rules.
Last, in Section \ref{sec:discussion}, we summaries the paper
and discuss future plans.

\section{Methods}\label{sec:methods}

\subsection{Multiple changepoint configurations}
Suppose we observe a time series data $\mathbf{X}_{1:n} = (X_1, X_2, \ldots, X_n)'$,
where $X_t$ is the observed value at time $t$. 
In multiple changepoint detection setting, there may exist multiple 
changepoints among time $\{1, \ldots, n\}$. 
More specifically, each of the time points (except the very few in the beginning, to avoid edge effects in 
time series modeling) can be either a changepoint or not, thus
making the total number of different multiple changepoint configurations to be 
on a scale of $2^n$.
Here, we refer to each possible multiple changepoint configuration a model.

Our goal is to select the most likely model given the observed data, i.e., to perform model selection.
The objective function for the model selection will be derived based on 
the Bayesian Minimum Description Length (BMDL) framework \citep{Li_etal_2019}. 
For each candidate model, we compute its BMDL score; and among all models we visit,
the one with the smallest BMDL score is the most optimal. Hence we select that model,
and declare all changepoints contained in that model to be the detected changepoints.

Suppose a multiple changepoint configuration (i.e., a candidate model) contains $m$ changepoints:
\[
1\leq \tau_1 < \tau_2 < \cdots < \tau_m \leq n.
\]
For notation simplicity, we denote $\tau_0 = 1$ and $\tau_{m+1} = n + 1$.
These changepoints partition the timeline into $m+1$ distinct  segments (i.e., regimes),
satisfying
\[
\text{Time } t \text{ is in regime } r \Longleftrightarrow \tau_{r-1} \leq t < \tau_r.
\]
As time progresses, the data experience some type of change when it hits a changepoint.
In this paper, the types of changes we consider include both mean shifts and slope changes. 
In other words, data in different regimes can have different mean levels and  linear slopes.

To better align with \citet{Li_etal_2019}, we adopt the same parameterization of the 
multiple changepoint configurations. 
For a candidate multiple changepoint model, instead of denoting it by $(m; \tau_1, \ldots, \tau_m)$,
whose length varied across models,  
we denote it by a fixed-length indicator vector $\boldsymbol{\eta} = (\eta_1, \ldots, \eta_n)$
such that
\[
\eta_t = 
  \begin{cases}
  1, & \text{ if time $t$ is a changepoint,}\\
  0, & \text{ if time $t$ is not a changepoint.}
  \end{cases}
\]
Then the total number of changepoints in that model can be recovered as $m = \sum_{t = 1}^n \eta_t$.

\subsection{Sampling distribution}
Under a given multiple changepoint model $\boldsymbol\eta$,
we assume that the observed data $X_t$ at time $t$ satisfies 
\begin{align}\nonumber
X_t = & \underbrace{\alpha_1 ~+~ \beta_1\cdot t}_{\color{blue}{\text{ linear segment in Regime 1(baseline)}}}
		+ \underbrace{\alpha_{r(t)} ~+~ \beta_{r(t)} \cdot t}_{\color{blue}{\text{increment linear segment in Regime } r}}\\ \label{eq:likelihood}
	& \underbrace{\sum_{i=1}^k \left[\theta_{i, 1} \cdot \sin\left(\frac{2\pi  t i}{T}\right)  
	+ \theta_{i, 2} \cdot \cos\left(\frac{2\pi  t i}{T}\right)\right]}_{
	\color{red}{\text{harmonic regression for the seasonal cycle}}}
		+ \underbrace{\epsilon_t}_{\text{AR}(p) \text{errors}}
\end{align}
Thus, we decompose the observed value $X_t$ into three additive components:
a regime-wise linear segment, a seasonal mean cycle, and an autocorrelated error.
Among these three, only the linear segment takes different values across different changepoint regimes,
while the other two components are global, in the sense that they 
do not experience changes overtime.

In the regime-wise linear segment, the intercept and slope parameters
differ across different regimes. 
In Regime 1, the intercept is $\alpha_1$ and the slope is $\beta_1$.
For any Regime $r$ where $r > 1$, the intercept and slope become
$\alpha_1 + \alpha_r$ and $\beta_1 + \beta_r$, respectively.
All these $\alpha$'s and $\beta$'s are treated as unknown parameters,
and will be estimated during the changepoint detection process. 
Since every changepoint model contains at least one regime, 
the baseline parameter pair $(\alpha_1, \beta_1)$ is included in all models,
although its estimated value varies under different models. 
On the other hand, the regime-wise incremental parameter pair 
$(\alpha_r, \beta_r)$ is model specific, in the sense that a model with
$m$ changepoints contains $m$ number of such pairs, with the subscript $r$
ranging from $2$ to $m+1$.

To incorporate a seasonal mean cycle, we include a harmonic regression \citep{Brockwell_Davis_2016} type of 
linear predictors in \eqref{eq:likelihood}. Here, $T$ is the period. 
For example, to reflect annual cycle in monthly data, we let $T = 12$;
while weekly cycle in daily data, $T = 7$. 
Since our goal is to automate the process of changepoint detection among many variables,
although all variables are collected at the same frequency, e.g., all monthly data, some of them may exhibit
seasonality while others may not. Furthermore, among those with seasonality,
the patterns of the seasonal mean cycles may be simply sinusoidal for some of them, 
while more complicated for the rest.
Therefore, we allow the harmonic regression order $k$ to vary by treating it as an unknown parameter.
It can take integer value from $0$ to a fixed upper bound $k_{\max}$.
To avoid singularity in the linear model \eqref{eq:likelihood}, 
we let $k_{\max} = \lfloor (T-1)/2 \rfloor$. For example, $k_{\max} = 5$ for monthly data.
The parameter $k$ determines the total number of 
harmonic regression coefficients $\theta_{1, 1}, \ldots, \theta_{k, 1}, ~  \theta_{1, 2}, \ldots, \theta_{k, 2}$
included in \eqref{eq:likelihood}. Similar to $\alpha$'s and $\beta$'s, these $\theta$'s are also 
unknown parameter to be estimated. 
Note that $k=0$ is equivalent to not including any seasonality components.

The last term in \eqref{eq:likelihood} is the error term. We let $\{\epsilon_t\}$ to be a
mean zero Gaussian autoregressive process of order $p$, i.e., 
\[
\epsilon_t = \sum_{j=1}^p \phi_j \epsilon_{t-j} +  Z_t, \quad Z_t \stackrel{iid}{\sim} \text{N}(0, \sigma^2)
\]
Here, the AR coefficients $\boldsymbol\phi = (\phi_1, \ldots, \phi_p)$ and the white
noise variance $\sigma^2$ are unknown parameters.
To accommodate both variables with independent errors and variables with autocorrelated errors, 
we permit the AR order $p$ to vary, from 0 to a fixed upper bound $p_{\max}$. 
Note that $p=0$ means independent errors. To avoid edge effects, we assume that there
are no changepoints among the first $p_{\max}$ times, i.e., we fix $\eta_t = 0$ for $t = 1, \ldots, p_{\max}$.

\subsection{BMDL expression}

Now we will derive the model selection objective function, the BMDL, for each
candidate model. Since the harmonic regression order $k$ and the AR order $p$ are allowed to vary, 
we generalize the concept of a candidate model, from a multiple changepoint configuration $\boldsymbol\eta$ alone,
to a combination of $\boldsymbol\eta$, $k$, and $p$.

We plan to follow the general guidelines of BMDL derivation, depicted in Section 3.1, 3.2 of
\citet{Li_etal_2019}.
To be consistent with the notation there, we denote the global coefficient vector
\[
\mathbf{s} = (\alpha_1, \beta_1)
\]
and trans-dimensional model specific coefficient vector 
\[
\boldsymbol\mu = (\alpha_2, \beta_2, \ldots, \alpha_{m+1}, \beta_{m+1},
\theta_{1, 1},  \theta_{1, 2}, \cdots, \theta_{k, 1}, \theta_{k, 2}).
\]
Among all model specific parameters, we apply mixture MDL to $\boldsymbol\mu$
under the independent normal prior distribution
\[
\boldsymbol{\mu} \sim \text{N}(\mathbf{0}, \nu \sigma^2 \mathbf{I}_{2m + 2k}),
\]
and apply two-part MDL to the rest of the parameters 
$\mathbf{s}, \sigma^2, \boldsymbol{\phi}$.
To compute additional penalty on the model, we let $\boldsymbol\eta$, $k$, and $p$
to have independent prior distributions as follows
\begin{align*}
\eta_t & \sim \text{Bernoulli}(\rho), t = p_{\max} + 1, \ldots, n, \quad \text{where } \rho \sim \text{Beta}(a, b),\\
k & \sim \text{Uniform}(0, 1, \ldots, k_{\max}),\\
p & \sim \text{Uniform}(0, 1, \ldots, p_{\max}).
\end{align*}

We omit the detail of BMDL derivation and discussions of hyper-parameter choices in this paper.
Interested readers can refer to the original BMDL paper \citep{Li_etal_2019} for more details. 
For a given model $(\boldsymbol\eta, k, p)$, its BMDL is 
\begin{align} \nonumber
\text{BMDL}(\boldsymbol\eta, k, p) 
= &~ \underbrace{\frac{n-p_{\max}}{2}\log \left( \hat{\sigma}^2 \right)}_{{\color{black}\text{goodness of fit}}} \\ \nonumber
&~	+ \underbrace{\frac{2m+2k}{2}\log(\nu) 
	+ \frac{1}{2}\log\left( \left|\widehat{\mathbf{D}}' 
	\widehat{\mathbf{D}}+\frac{\mathbf{I}_{2m+2k}}{\nu}\right| \right)}_{{\color{black}\text{penalty on $\boldsymbol\mu$, for linear segments and seasonality coefficients}}}\\ \nonumber
&~	+ \underbrace{\frac{p}{2}\log (n-p_{\max})}_{{\color{black}\text{penalty on the AR coefficient $\boldsymbol\phi$}}}\\ \label{eq:bmdl}
&	\underbrace{- \log\left[\Gamma\left(a + m\right) \Gamma\left(b+ n-p_{\max}-m\right) \right]}_{{\color{black}\text{penalty on $\boldsymbol\eta$; 
	increases with $m$}}},
\end{align}
where $\Gamma(\cdot)$ is the Gamma function, and terms such as 
$\hat{\sigma}^2$ and $\widehat{\mathbf{D}}$ are introduced
during the BMDL derivation (see \citep{Li_etal_2019} for their definitions). 
Thus, the BMDL method can be viewed as a penalization approach, where the 
objective function balances the goodness of fit and the complexity of the model.

\subsection{Computation: Markov chain Monte Carlo}

Since there are $(2^{n-p_{\max}}) \times k_{\max} \times p_{\max}$ number of distinct
candidate models in total, even with the fastest computer in the world, it is impossible to visit each and every
model. To quickly explore the part of model space
that contains good models, we can use stochastic model search algorithms. 
Similar to the computation strategy adopted in \citep{Li_etal_2019}, here we 
use a specific type of MCMC, the Metropolis-Hastings algorithm, 
which takes turns to update $\boldsymbol\eta$, $k$ and $p$, one at a time. 
According to our empirical experience, usually an optimal model can be found
within a reasonable number of MCMC iterations (say $10^5$ for a five-year-long monthly data series).

\section{Results}\label{sec:results}
\subsection{Average hourly pay in metropolitan areas}

To demonstrate the performance of our method in a static offline manner, 
we apply it to a dataset on hourly pay in metropolitan areas.
This dataset is provided by Bureau of Labor Statistics, and is publicly available online (\url{https://www.bls.gov/sae}).
For a metropolitan area, Bureau of Labor Statistics estimates a monthly time series
of average hourly earnings of workers on non-farm payrolls in that area, based on their surveys.
The time series dataset we use here spans from January 2011 to October 2017.
We study the  earnings for several large cities, by independently applying
the BMDL approach described above for each of them in parallel. 

\begin{figure}[h]
  \centering
  \includegraphics[width=\linewidth]{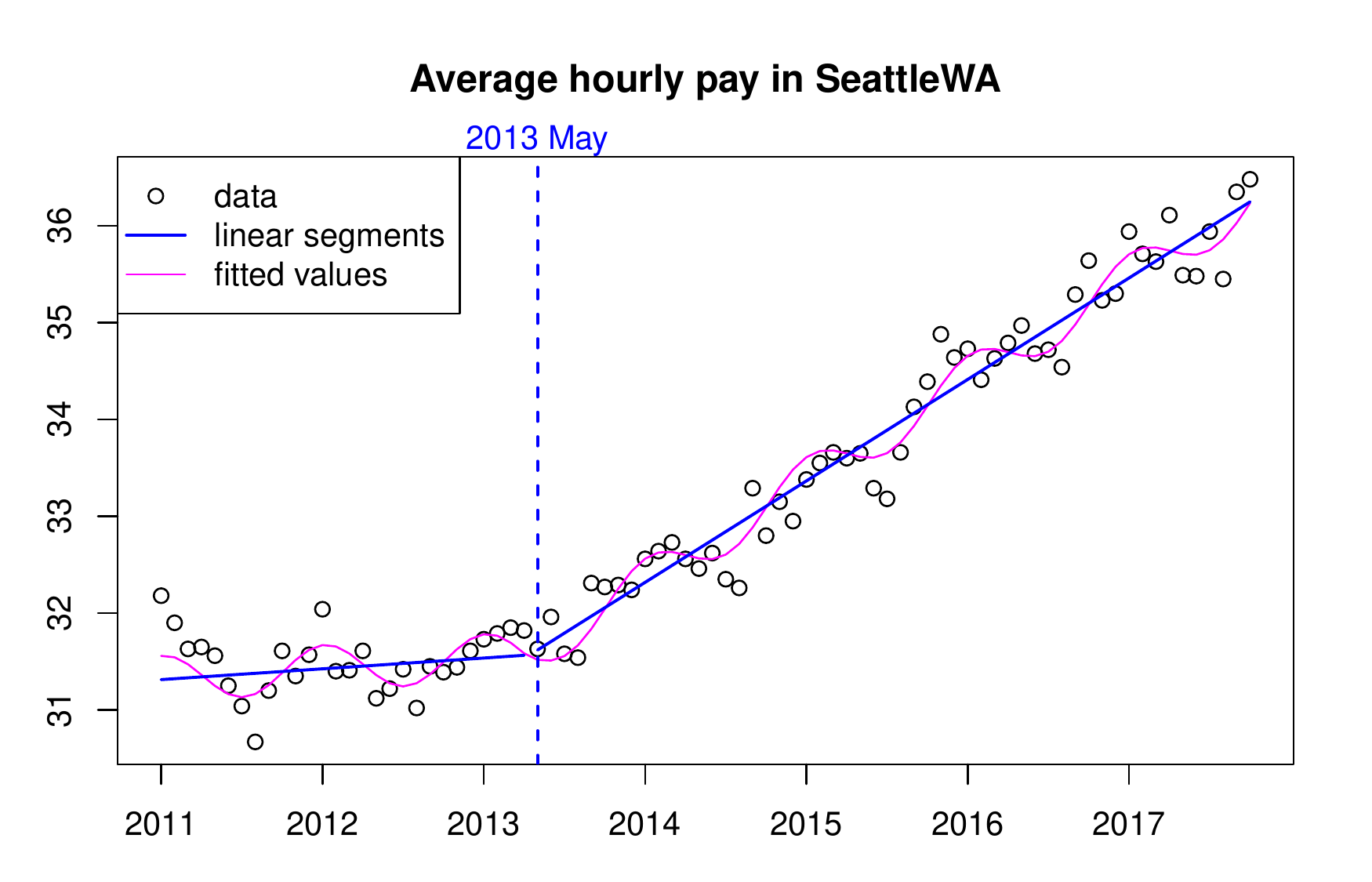}
  \caption{Hourly pay in Seattle WA: a changepoint in May 2013 is detected,
  with a seasonal mean cycle and independent errors. 
  The estimated linear segment is $31.2981 + 0.0093 t$ in Regime 1 
  and $29.0806 + 0.0873 t$ in Regime 2.}\label{fg:Seattle}
\end{figure}

Let us first take Seattle WA as an example. In Figure \ref{fg:Seattle}, 
the x-axis is time and y-axis is hourly pay in US dollars. 
The dots are the observed data; 
the dashed vertical line(s) are the detected changepoint(s).
For Seattle, one changepoint in May 2013 is detected, which separate
the timeline into two regimes. 
The blue lines indicate the regime-wise linear segments.
Under the detected multiple changepoint model, we obtain estimates
on the intercepts and slopes of these linear segments.
Before the May 2013 changepoint, Seattle workers' average earning increases
at a very mild rate (estimated slope $0.009$), almost stays the same, while after the changepoint, 
it increases at a much faster speed  (estimated slope $0.087$). 
The purple line indicates the fitted value of the linear model \eqref{eq:likelihood}, 
i.e., the sum of the linear segment and the seasonal mean cycle. 
Apparently, in addition to the linear trend, 
average pay in Seattle also experiences a sinusoidal shaped annual cycle.
Here, our BMDL method selects $k=1$ for the harmonic regression order 
and $p=0$ for the AR order of errors. 

\begin{figure}[h]
  \centering
  \includegraphics[width=\linewidth]{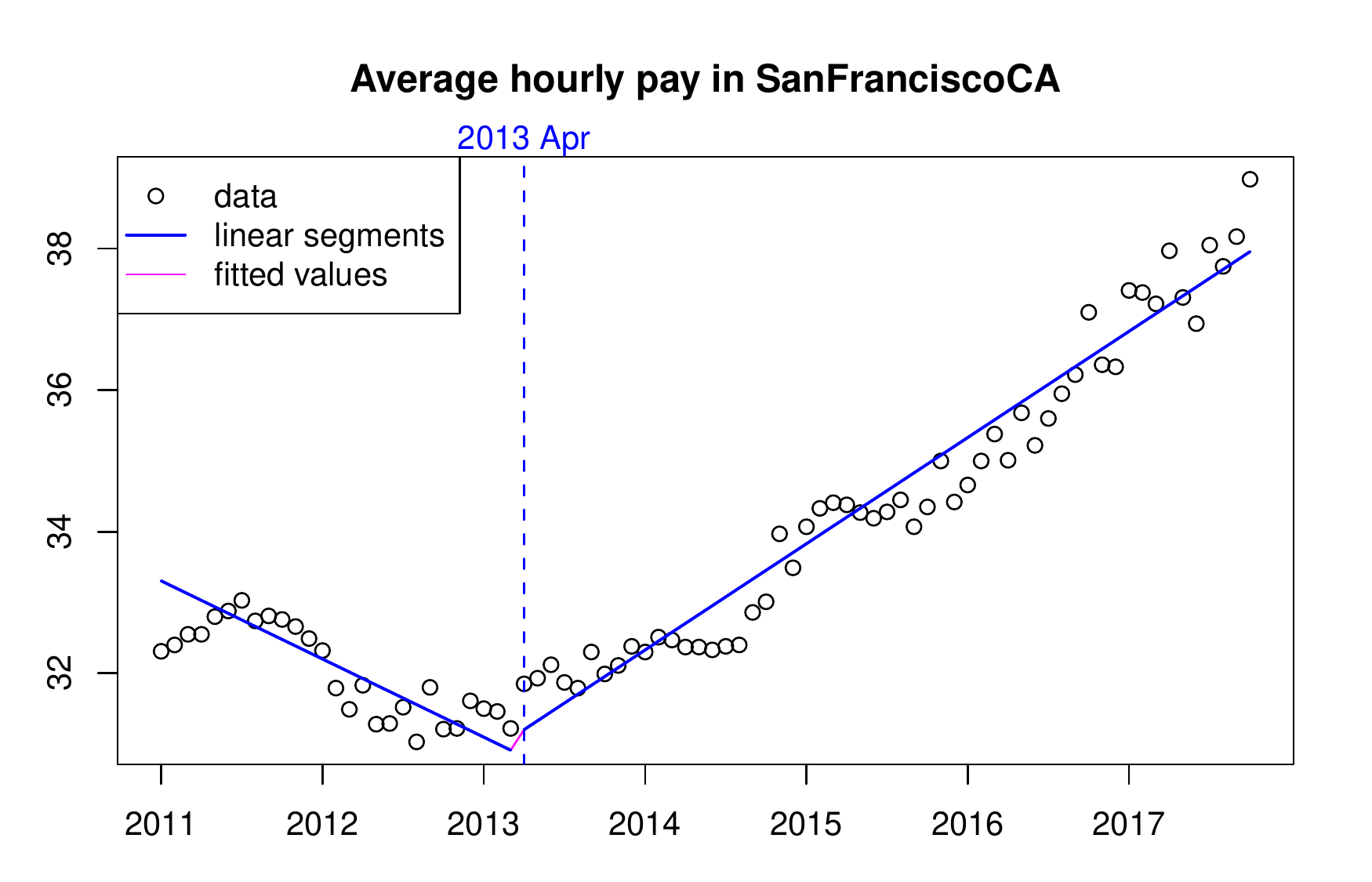}
  \caption{Hourly pay in San Francisco CA: a changepoint in April 2013 is detected,
  with AR(2) errors but no seasonal mean cycle. 
  The estimated linear segment is $33.3974 - 0.0920 t$ in Regime 1 
  and $27.7010 + 0.1251 t$ in Regime 2.}\label{fg:SanFransisco}
\end{figure}

For San Francisco CA, we detect one changepoint in April 2013 (see Figure \ref{fg:SanFransisco}). 
Interestingly, the average earning in San Fransisco first decreases overtime 
before early 2013 (estimated slope $-0.092$), and then increases at a very high rate
(estimated slope $0.125$), even faster than Seattle's 
pace in its Regime 2. 
Unlike Seattle, here no seasonal cycles are detected (so that the purple line
overlaps with the blue line).
Moreover, the estimated error series is no longer independent;
rather, it is AR with order $p=2$. The estimates of the AR coefficients are $\phi_1 = 0.4099, \phi_2 = 0.2139$.

\begin{figure}[h]
  \centering
  \includegraphics[width=\linewidth]{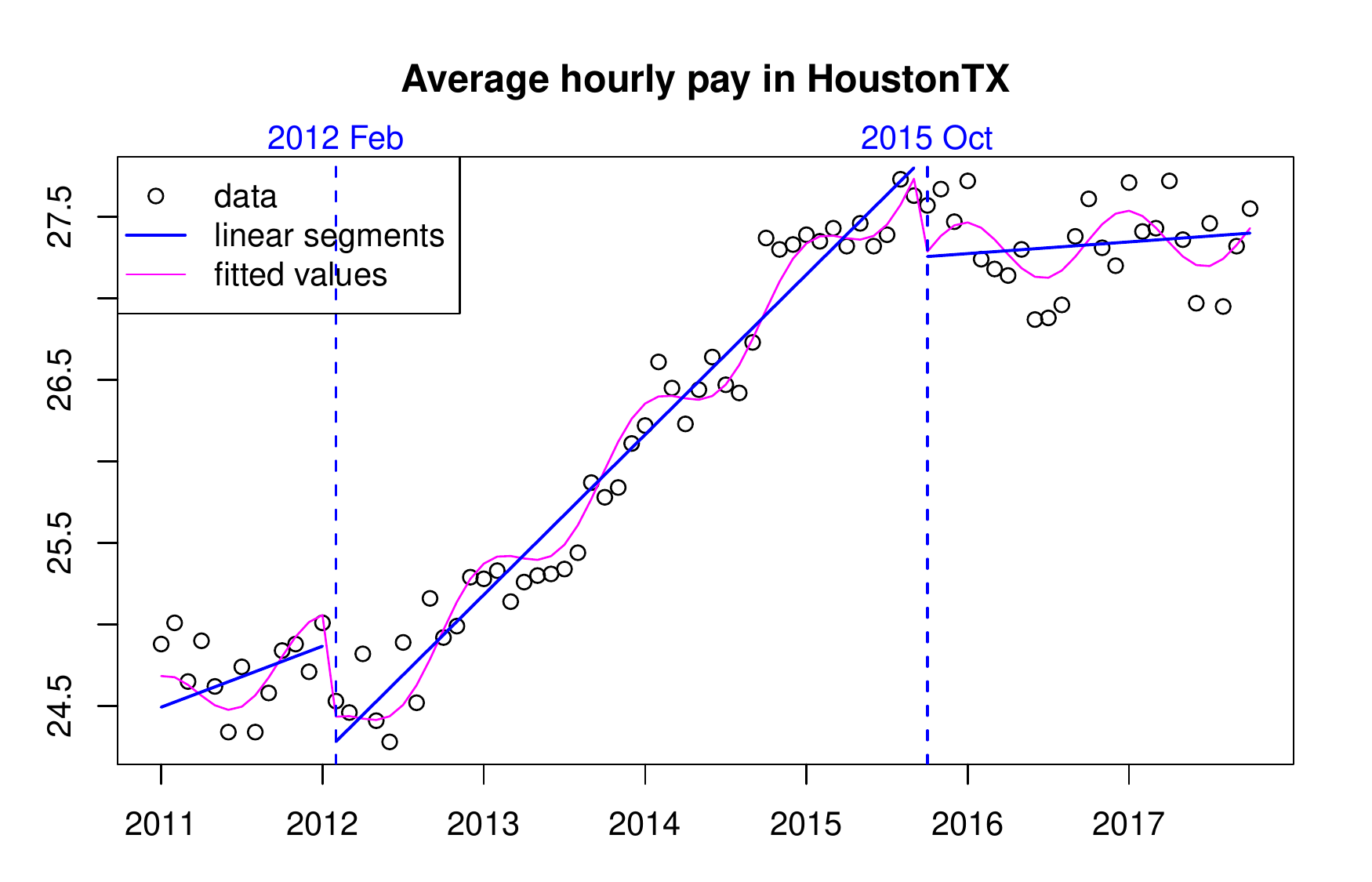}
  \caption{Hourly pay in Houston TX: two changepoints are detected.}\label{fg:Houston}
\end{figure}

For both Seattle and San Fransisco, the selected multiple changepoint models
contain exactly one changepoint. For some other cities, the optimal 
candidate model our algorithm detects contains more than one changepoints. 
For example, we detect two changepoints for Houston TX, one in February 2012, 
and the other in October 2015 (see Figure \ref{fg:Houston}). 

\begin{figure}[h]
  \centering
  \includegraphics[width=\linewidth]{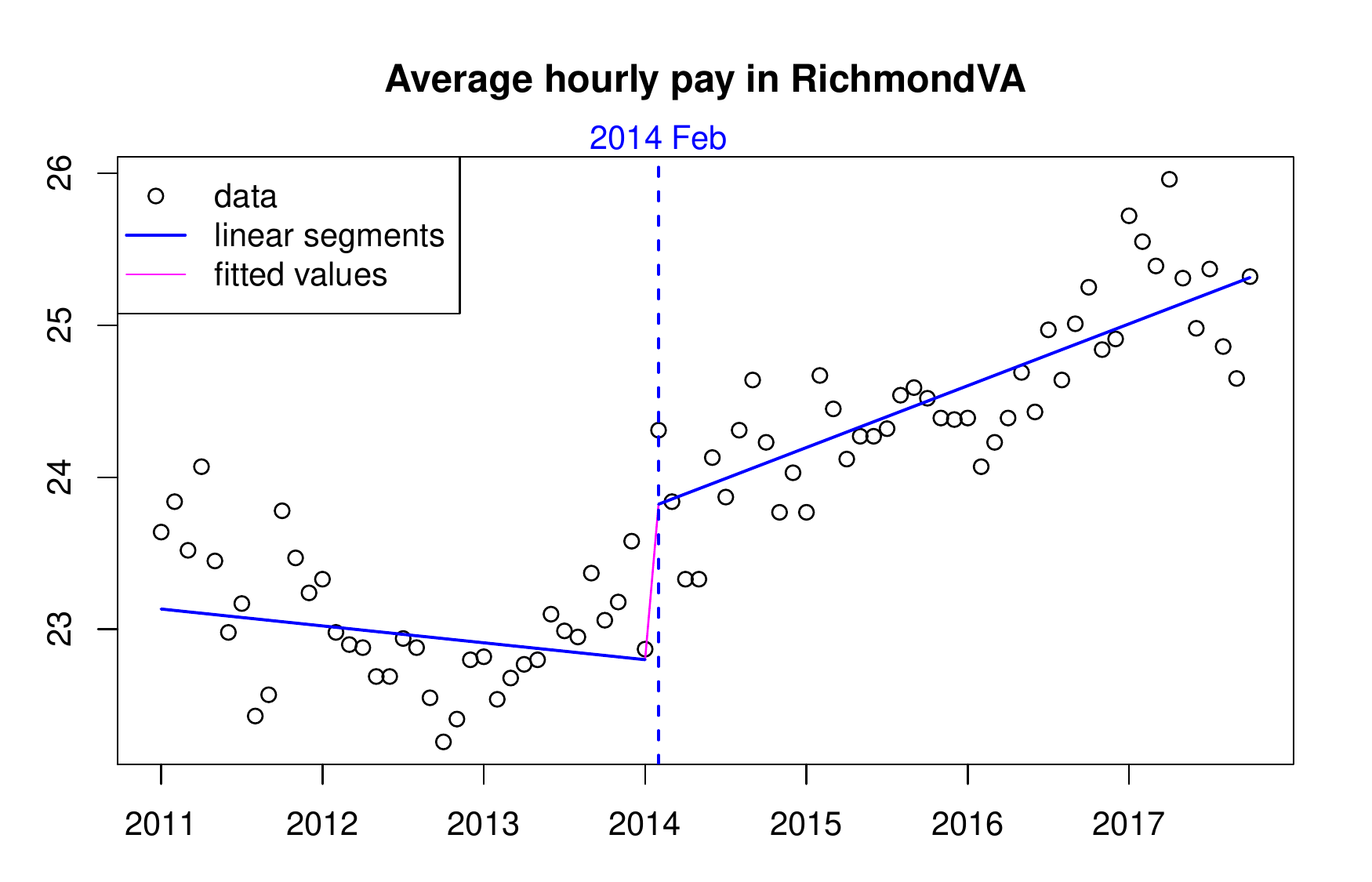}
  \includegraphics[width=\linewidth]{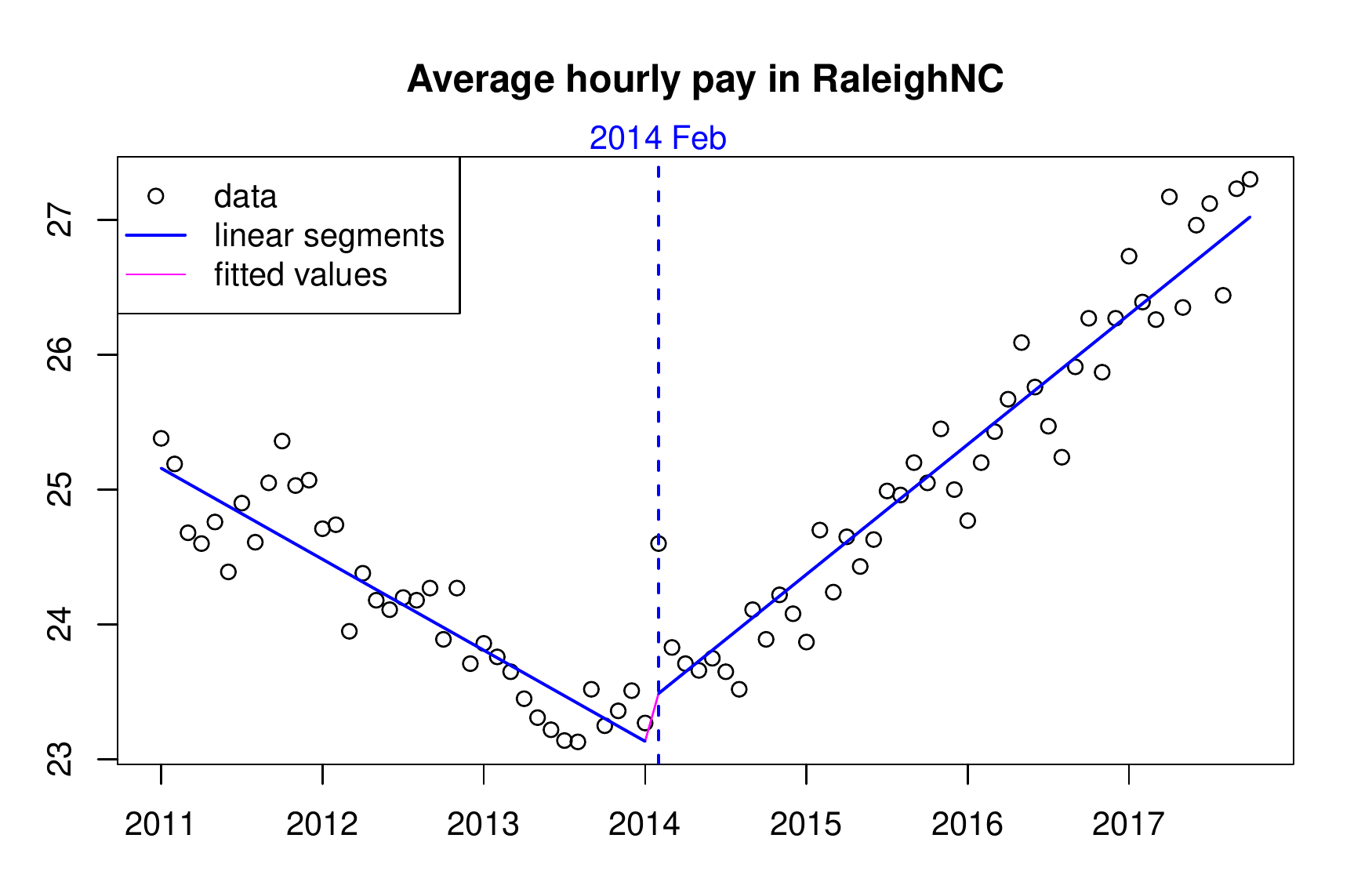}
  \includegraphics[width=\linewidth]{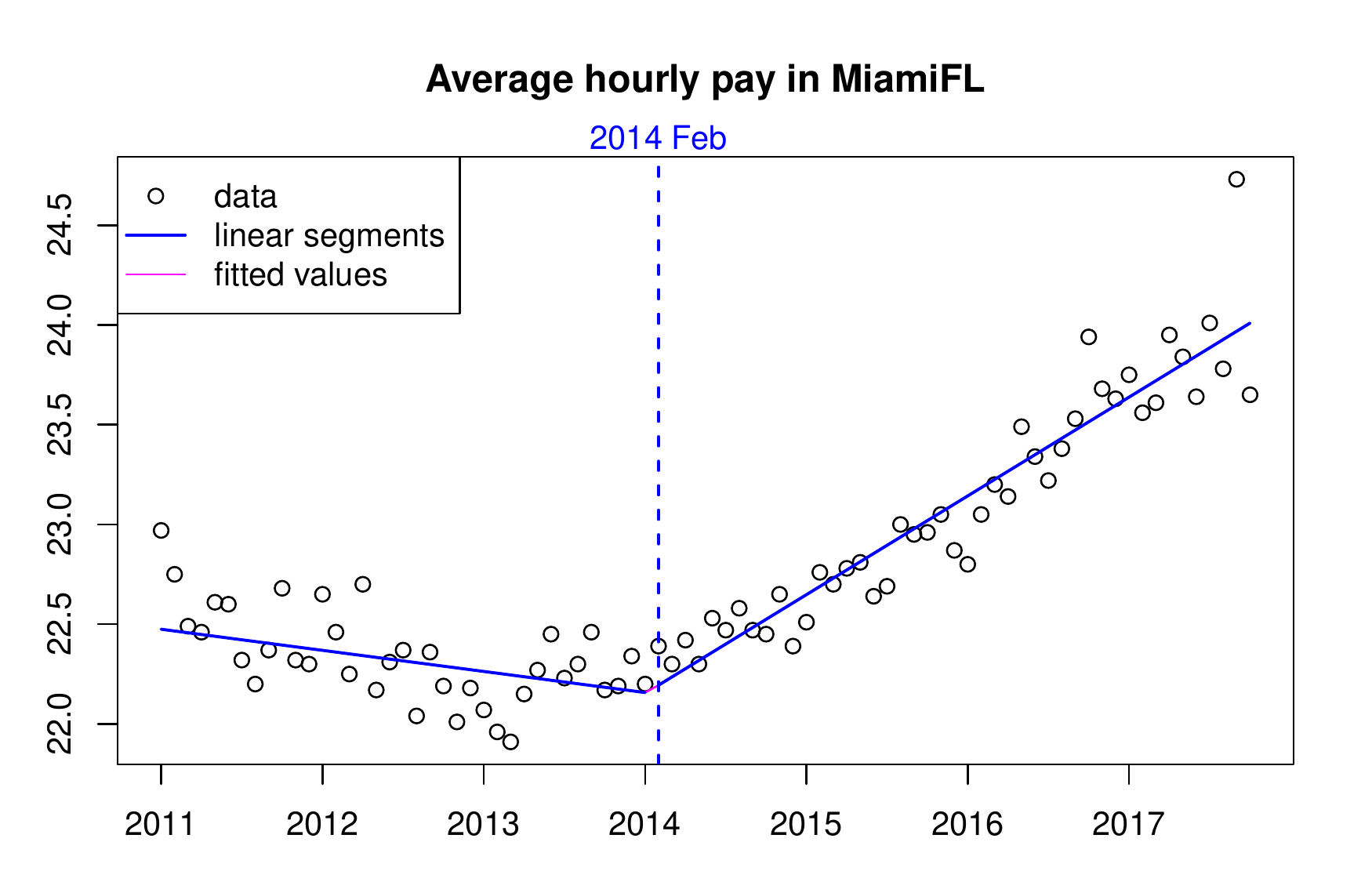}
  \caption{Hourly pay in Richmond VA, Raleigh NC, and Miami FL (from top to bottom) have
  a common changepoint in Febuary 2014.}\label{fg:southeast}
\end{figure}

One interesting phenomenon is that the detected changepoints for Seattle and San Fransisco 
are very close to each other -- just one month apart. 
This may not be a coincidence. After all, these two large cities on the west coast 
have similar types of industries, as they are the home of many rapidly growing tech companies. 
We also notice a similar pattern among large cities on the southeast. 
For example, a common changepoint in February 2014
is detected for Richmond VA, Raleigh NC, and Miami FL (see Figure \ref{fg:southeast}).

\subsection{A simulation example}

Data monitoring is often an on-going task rather than a one time effort, as new data keep coming in
every month. Thus, rather than detecting all historical changes in the past,
the goal is to detect the most recent change as soon as it occurs. 
To perform data monitoring in this online manner, 
we repeatedly apply our BMDL changepoint method every month as new data points are included.  
Each time, we use all historical data, along with the data point newly come in, as the input 
to the BMDL algorithm. 
In this subsection, we use simulation data to demonstrate the performance of BMDL in data monitoring, 
and compare it with the traditional control chart based method. 

To thoroughly study true positives and false positives, 
we simulate data under various scenarios. 
Under each scenario, we independently generate 100 time series realizations.
Each simulated time series contains $n = 500$ data points and satisfies the
assumption as being the sum of three components: a linear segment, 
a seasonal mean cycle, and an AR(1) error.
For the AR error, we fix the white noise variance $\sigma^2 = 1$, 
and use different values for the AR coefficient $\phi = 0.3, 0.5, 0.7$
to represent different levels of autocorrelation, from mild to high. 
For the seasonality, we use a fixed fluctuation pattern whose range equals 6.
For the linear segment, we only put one changepoint at time $t=60$, and 
explore different combinations of slope changes and mean level jumps at the changepoint, 
according to Table \ref{tb:simulation_setup}.
Hence, there are 4 scenarios with no changes, and $5 \times 21 - 1 = 104$ scenarios with 
one changepoint. (Here the minus one is because the scenario with zero trend change and 
zero jump is not a scenario with a changepoint.)

\begin{table}
\caption{Simulation setup for the linear segments.}\label{tb:simulation_setup}
\begin{tabular}{| c | c | c|}
\hline
Scenarios						& Trend slope					&  Jump				\\ \hline
							& 0 	$\rightarrow$ 0			& \multirow{4}{*}{0}		\\ \cline{2-2}
No							& 0.1 $\rightarrow$ 0.1		& 					\\ \cline{2-2}
change						& 0.2 $\rightarrow$ 0.2		& 					\\ \cline{2-2}
							& 0.3 $\rightarrow$ 0.3		& 					\\ \hline
\multirow{5}{*}{Change} 			& 0 	$\rightarrow$ 0			& 10,					\\ \cline{2-2}
							& 0.05 $\rightarrow$ -0.05		&  9,					\\ \cline{2-2}
							& 0.1 $\rightarrow$ -0.1		& $\vdots$			\\ \cline{2-2}
							& 0.2 $\rightarrow$ -0.2		& -9,					\\ \cline{2-2}
							& 0.3 $\rightarrow$ -0.3		& -10				\\ \hline
\end{tabular}
\end{table}

For each time series, starting from time $t = 60$, we apply 
the BMDL in a repeatedly online manner with one additional data point at a time.
Once a changepoint is detected, either at time 60 exactly or at another location, 
we stop and declare detection for this series.
On the other hand, if at the end when the whole
series of length $n=500$ is exhausted and there are still no changepoints detected, 
then we declare a no detection for this series.

The control chart approach we compare with is Shewhart rules \citep{Shewhart_1931}.
In particular, an alert is triggered if 
 \begin{itemize}
  \item 1 most recent point is beyond 4 sigma, or
  \item 2 out of most recent 3 points are  beyond 3 sigma and on the same side of the centerline, or
  \item 8 most recent points are beyond 1 sigma and on the same side of the centerline.
  \end{itemize}
Once any one of these rules is triggered, we stop and and declare a detection. 
Here, sigma is the sample standard deviation among the data in the benchmark period. 
Ideally, the benchmark period should be changepoint free, and the samples
within it should be independently and identically distributed (iid).  
Since we are aware of the underlying truth for this simulation study, we set the benchmark to be the period 
before the changepoint, i.e., from time 1 to time 59. 
Note that this is not a feasible solution in real world since the underlying data structure is not revealed to us.

\begin{figure}[h]
  \centering
  \includegraphics[width=\linewidth]{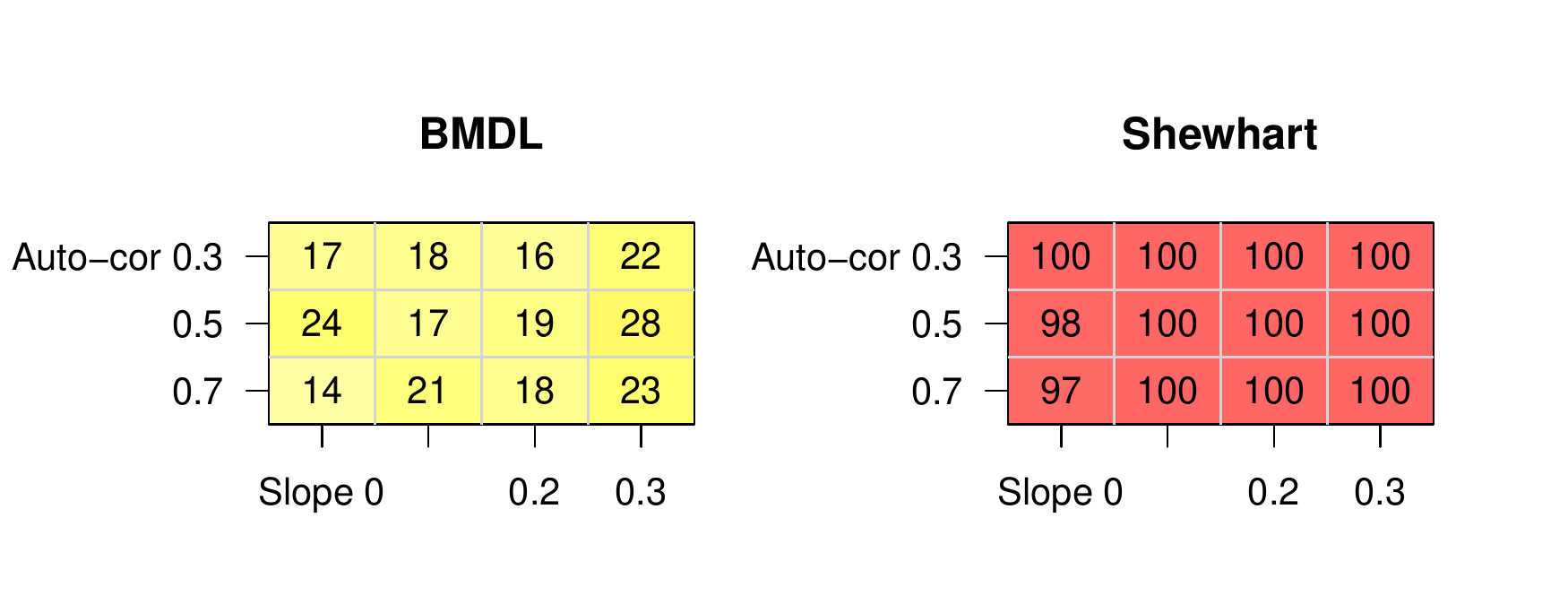}
  \caption{Detection rates for simulation data, false positives.}\label{fg:simulation_FP_rate}
\end{figure}

We first examine the false positive detection rates. 
Figure \ref{fg:simulation_FP_rate} summarizes the number of realizations (out of 100)
that we declare detection, under the scenarios where there are no actual changes. 
We find that
regardless of the linear slope or the autocorrelation, our BMDL method has
only about $20\%$ false positives, while the Shewhart rules have almost $100\%$!
This is not surprising, since Shewhart rules are supposed to work under iid assumption, 
i.e., no trend, seasonality, or autocorrelation, which is apparently not true for
the simulation data here (neither does it hold for most real world applications).
Any of the nonzero slope, seasonal cycle, or autocorrelation in the data may trigger alerts,
thus leading to the extremely high possibility of false positives. 

As to the true positives, Figure \ref{fg:simulation_FP_rate} indicates that BMDL and the 
Shewhart rules perform equally well most of the time.
For all scenarios except where the signal-to-noise level is too small (no slope change, jump size $1$), 
their detection rates are both $100\%$. 
However, when the signal is very weak, BMDL only detects changes about $50\%\sim 60\%$ of the time,
while Shewhart rules still have more than $90\%$ detection rates.
This is because Shewhart control chart rules tend to trigger alert easily
(recall the $100\%$ false positive rates), while penalization based methods such as 
the BMDL \eqref{eq:bmdl} often prefer parsimonious models, unless there exist
strong evidence to favor more complicated models. 

\begin{figure}[h]
  \centering
  \includegraphics[width=\linewidth]{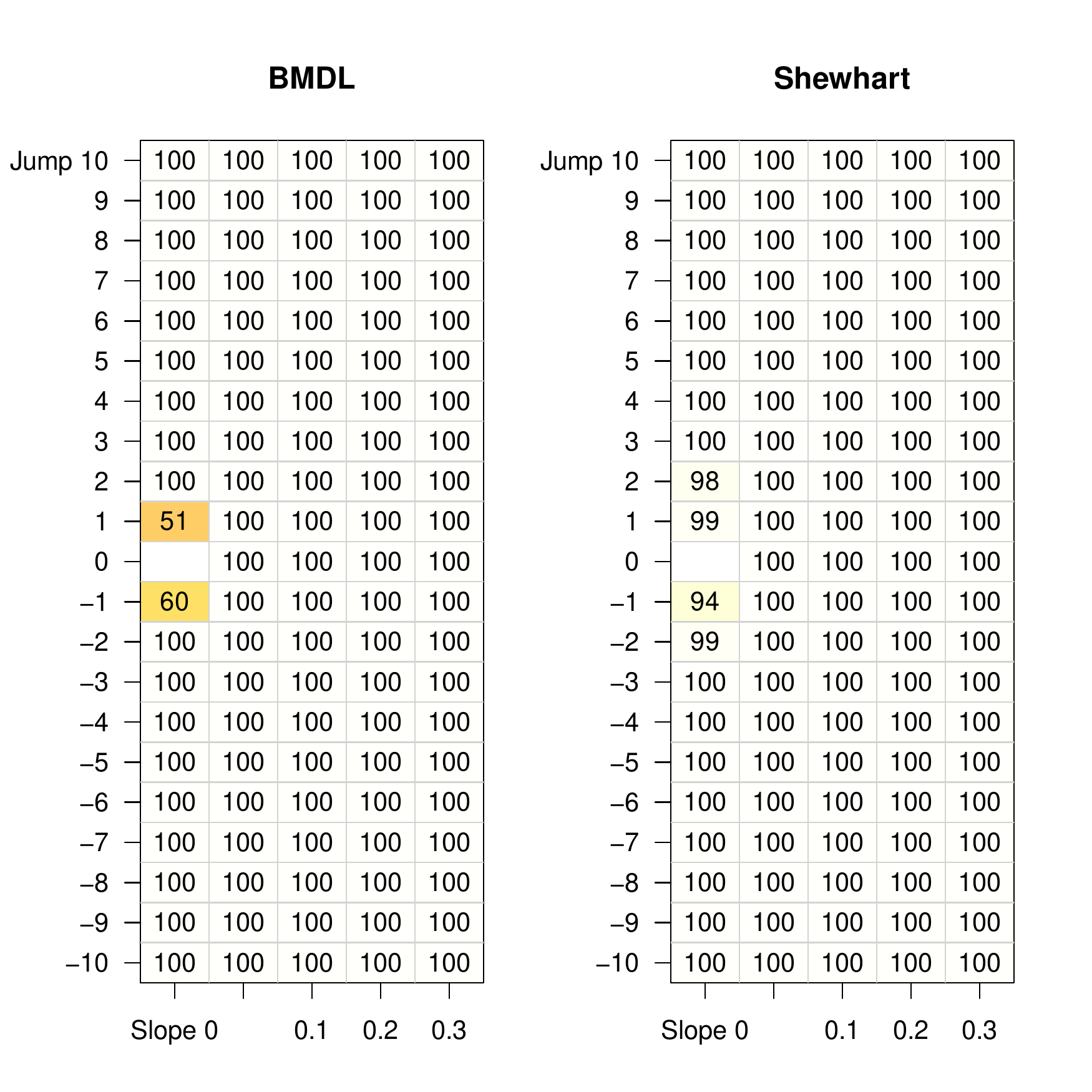}
  \caption{Detection rates for simulation data, true positives, $\phi = 0.3$.}\label{fg:simulation_TP_rate}
\end{figure}

\begin{figure}[h]
  \centering
  \includegraphics[width=\linewidth]{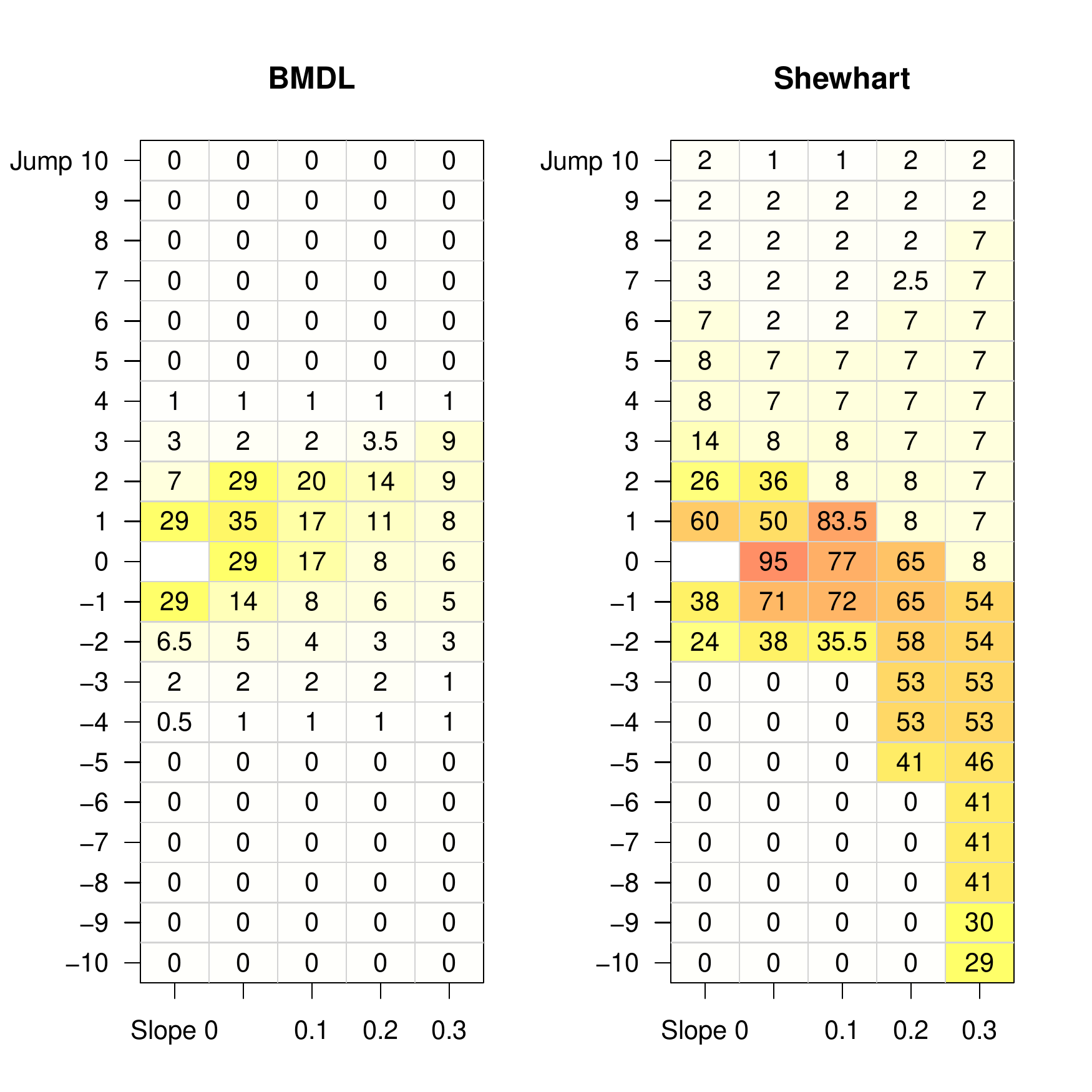}
  \caption{Median run length among true positives, $\phi = 0.3$.}\label{fg:simulation_TP_speed}
\end{figure}

In addition to detection, it is also important to detect the changes as quickly as possible. 
Here, we call the run length to be the difference between time 60 and the time of detection. 
For example, the run length is 0 if we declare detection when the last data point included is time 60, and 1 if
we declare detection when the last data point is time 61.
Figure \ref{fg:simulation_TP_speed} show the median run length among detected 
 realizations. 
BMDL method systematically outperforms Shewhart rules. In particular,
when signals are strong, for example, for scenarios with jump size 5 or greater, 
BMDL is always able to detect changes right on spot, i.e., as soon as when time 60 is included.
On the contrary, detection of the Shewhart rules can be 7 or more time points slower.

\section{Discussion}\label{sec:discussion}

In this paper, we have introduced a novel extension of the 
BMDL multiple changepoint detection method. 
It detects changepoints in terms of both mean level shifts and linear slopes,
and also whether the data have seasonality or autocorrelation.
It enables explainable anomaly detection by providing
estimations on the intercepts and slopes for regime-wise linear segments,
and harmonic regression coefficients for the seasonality component. 
Our simulation example illustrates that compared with
traditional control chart tools, our method substantially
reduces false positive alarms. In addition, it performs almost equally well among
true positives, and usually detects anomalies in a timely manner. 

As to future plans, various directions can be explored. 
One possibility is to extend this method to multivariate time series. 
While this paper is about detecting multiple changepoints in a univariate time series,
in real business applications, there are usually correlations across multiple series.
In fact, changepoint detection on multivariate data has been receiving significant 
attention in recent years (see \citet{Li_etal_2019} and the references therein).
We can model these correlations in the sampling distribution 
by replacing the autoregressive process by the vector autoregressive process (VAR).
Another extension is to apply to the BMDL framework to different
types of time series processes, such as ARMA and GARCH.

%
\bibliographystyle{ACM-Reference-Format}
\bibliography{sample-base}


\begin{thebibliography}{3}


\ifx \showCODEN    \undefined \def \showCODEN     #1{\unskip}     \fi
\ifx \showDOI      \undefined \def \showDOI       #1{#1}\fi
\ifx \showISBNx    \undefined \def \showISBNx     #1{\unskip}     \fi
\ifx \showISBNxiii \undefined \def \showISBNxiii  #1{\unskip}     \fi
\ifx \showISSN     \undefined \def \showISSN      #1{\unskip}     \fi
\ifx \showLCCN     \undefined \def \showLCCN      #1{\unskip}     \fi
\ifx \shownote     \undefined \def \shownote      #1{#1}          \fi
\ifx \showarticletitle \undefined \def \showarticletitle #1{#1}   \fi
\ifx \showURL      \undefined \def \showURL       {\relax}        \fi
\providecommand\bibfield[2]{#2}
\providecommand\bibinfo[2]{#2}
\providecommand\natexlab[1]{#1}
\providecommand\showeprint[2][]{arXiv:#2}

\bibitem[\protect\citeauthoryear{Brockwell and Davis}{Brockwell and
  Davis}{2016}]%
        {Brockwell_Davis_2016}
\bibfield{author}{\bibinfo{person}{Peter~J. Brockwell} {and}
  \bibinfo{person}{Richard~A. Davis}.} \bibinfo{year}{2016}\natexlab{}.
\newblock \bibinfo{booktitle}{\emph{Introduction to Time Series and
  Forecasting, Third Edition}}.
\newblock \bibinfo{publisher}{New York: Springer}.
\newblock


\bibitem[\protect\citeauthoryear{Li, Lund, and Hewaarachchi}{Li
  et~al\mbox{.}}{2019}]%
        {Li_etal_2019}
\bibfield{author}{\bibinfo{person}{Yingbo Li}, \bibinfo{person}{Robert Lund},
  {and} \bibinfo{person}{Anuradha Hewaarachchi}.}
  \bibinfo{year}{2019}\natexlab{}.
\newblock \showarticletitle{Multiple Changepoint Detection with Partial
  Information on Changepoint Times}.
\newblock \bibinfo{journal}{\emph{Electronic Journal of Statistics, accepted}}
  (\bibinfo{year}{2019}).
\newblock
\urldef\tempurl%
\url{https://arxiv.org/abs/1511.07238}
\showURL{%
\tempurl}


\bibitem[\protect\citeauthoryear{Shewhart}{Shewhart}{1931}]%
        {Shewhart_1931}
\bibfield{author}{\bibinfo{person}{Walter~Andrew Shewhart}.}
  \bibinfo{year}{1931}\natexlab{}.
\newblock \bibinfo{booktitle}{\emph{Economic control of quality of manufactured
  product}}.
\newblock \bibinfo{publisher}{ASQ Quality Press}.
\newblock


\end{thebibliography}

\end{document}